\documentclass[aps,twocolumn,showpacs]{revtex4}

\usepackage{graphicx}

\begin{document}

\title{Statistical properties of a dissipative kicked system: critical
exponents and scaling invariance}

\author{Diego F. M. Oliveira$^{\rm 1}$}
\author{Marko Robnik$^{\rm 1}$}
\author{Edson D.\ Leonel$^{\rm 2}$}
\affiliation{$^{\rm 1}$CAMTP - Center For Applied Mathematics and
Theoretical Physics  University of Maribor - Krekova 2 - SI-2000 -
Maribor - Slovenia.\\
$^{\rm 2}$Departamento de Estat\'{\i}stica, Matem\'atica Aplicada e
Computa\c c\~ao - UNESP -- Univ Estadual Paulista -- Av. 24A,
1515 -- Bela Vista -- 13506-900 -- Rio Claro -- SP -- Brazil.}

\date{\today} \widetext


\begin{abstract}
A new universal {\it empirical} function that depends on a single
critical exponent (acceleration exponent) is proposed to describe the
scaling behavior in a dissipative kicked rotator. The scaling formalism
is used to describe two regimes of dissipation: (i) strong dissipation
and (ii) weak dissipation. For case (i) the model exhibits a route to
chaos known as period doubling and the Feigenbaum constant along the
bifurcations is obtained. When weak dissipation is considered the
average action as well as its standard deviation are described using
scaling arguments with critical exponents. The universal {\it empirical}
function describes remarkably well a phase transition from limited to
unlimited growth of the average action.
\end{abstract}

\maketitle

\section{Introduction}

In 1969 Boris Chirikov \cite{ref0,ref1} proposed what became one of
the most important and extensively studied systems in nonlinear dynamics
and in the theory of Hamiltonian systems and area-preserving maps
\cite{ref2,ref3}, namely the Chirikov standard map. The model describes
the motion of a kicked rotator. Applications of the model can be
made in different fields of science including solid state physics
\cite{ref6}, statistical mechanics \cite{ref8} and accelerator physics
\cite{ref4}. It has also been studied in relation to problems of quantum
mechanics and quantum chaos \cite{ref7,ref7a}, plasma physics
\cite{ref5} and many others.

The standard map is a dynamical system in which the nonlinearity given
by a sine function is controlled by a parameter $K$. In the absence of
dissipation, if $K$ is small enough the structure of the phase space is
mixed in the sense that Kolmogorov-Arnold-Moser (KAM) invariant tori and
islands are observed coexisting with chaotic seas \cite{ref9,ref10,ref11,ref12,ref13}.
 As the parameter $K$
increases and becomes larger than $K_c\approx 0.971635\cdots$, the last
invariant spanning curve disappears and the system presents a global
chaotic component where a chaotic orbit spreads over the phase space.
However, the introduction of dissipation in the model changes the mixed
structure and the system exhibits attractors \cite{ref131,ref132,ref133,ref1334}. We consider two different
ranges of dissipation namely: (i) strong dissipation -- the situation
where the variable action loses more than $70\%$ of its value upon
a kick -- where period doubling  bifurcation cascade is observed and the
Feigenbaum coefficient is numerically obtained and; (ii) weak
dissipation where the average action exhibits scaling features
\cite{ref13a}. Here we propose a new universal function
that describes the scaling behavior of the average action by the
knowledge of a single critical exponent, the so called acceleration
exponent.

The paper is organized as follows. In section \ref{sec2} we describe all
the necessary details to obtain the two-dimensional map that describes
the dynamics of the system and also we present and discuss our numerical results. 
Conclusions are drawn in Sec. \ref{sec4}.

\section{The model and results}
\label{sec2}

The Hamiltonian that describes the standard map has the  following form (see e.g., \cite{ref1,ref2,ref14}):
\begin{equation}
H={I^2 \over 2}+K\cos(\theta) \sum_{n=-\infty}^{\infty}\delta \left({t \over T} -n \right)
\label{eq1}
\end{equation}
where $K$ is the amplitude of the delta-function kicks (pulses) with period $T=2\pi/\nu$.
The equation of motion is given by
\begin{eqnarray}
\dot{I}=K\sin(\theta)\sum_{n=-\infty}^{\infty}\delta \left({t \over T} -n \right)~~,~~
\dot{\theta}=I~.
\label{eq2}
\end{eqnarray}
Assuming that $(I_n,\theta_n)$ be the values of the variable immediately after the $n^{th}$ kick, $(I_{n+1},\theta_{n+1})$ represent their values after
the $({n+1})^{th}$ kick, and introducing the dissipation parameter $\gamma$ \cite{ref14a}, the dissipative standard map is written as
\begin{eqnarray}
S:\left\{\begin{array}{ll}
I_{n+1} = (1-\gamma) I_n+K \sin(\theta_n)  \\
{\theta}_{n+1} = \theta_n+I_{n+1} ~~~~~~~ ~~~mod(2\pi) 
\end{array} ~,
\right.
\label{eq3}
\end{eqnarray}
where $ \gamma\in[0,1]$ is the dissipation parameter. If $\gamma=0$ all the results for the Hamiltonian area-preserving standard map are recovered. On the other hand,
if $\gamma=1$ the results for the one-dimensional sine-circle map are obtained \cite{ref2}. In this paper we shall consider $0<\gamma<1$. 
The system is area preserving only when $\gamma=0$ since the determinant of the Jacobian matrix is $\det(J)=1-\gamma$. Another consequence of
the dissipation is that the map (\ref{eq3}) is periodic only in $\theta$, while the action $I$ ranges between $-\infty <I<\infty$. 

\begin{figure}[t]
\centerline{\includegraphics[width=1.0\linewidth]{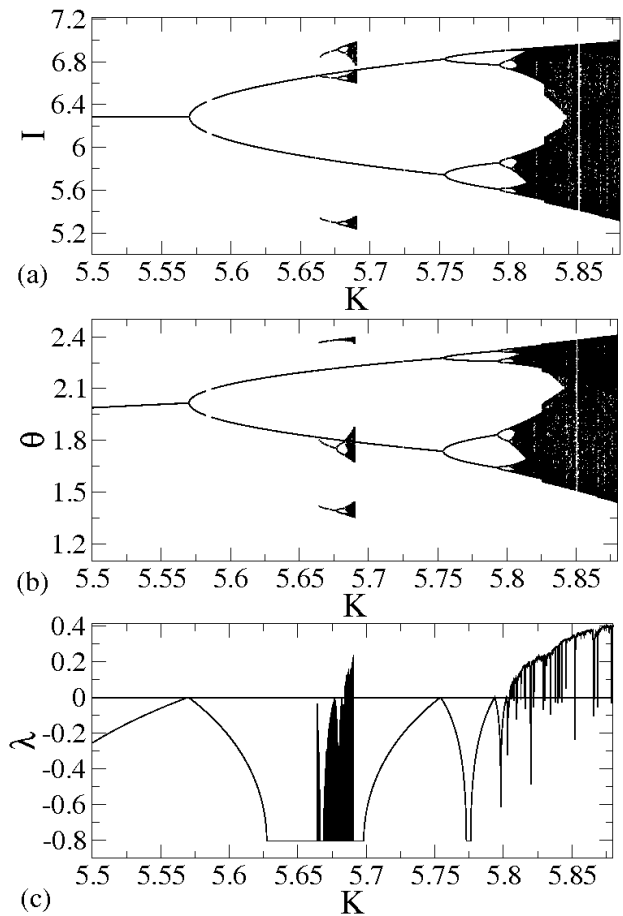}}
\caption{Bifurcation cascade for (a) $I$ and (b) $\theta$ both plotted
against the parameter $K$. In (c) it is shown the Lyapunov exponent associated
to (a) and (b). The damping coefficients used for the construction of
the figures (a), (b) and (c) were $\gamma=0.80$.}
\label{fig1}
\end{figure}

In this paper we will consider two situations for $\gamma$, namely, strong dissipation and weak dissipation.
The strong dissipation refers to  the situation in
which the action variable $I$ loses more than $70\%$ of its value upon a kick. We have considered the case $\gamma=0.80$. To explore some typical behavior we have used as initial conditions
$\theta_0=3$ and $I_0=5.53$ and investigated its attraction to periodic orbits, and looked at the bifurcations as $K$ varies in the range where we have global chaos if $\gamma=0$. Fig. \ref{fig1}(a) shows
the behavior of the asymptotic action plotted against the control
parameter $K$, where a sequence of period doubling bifurcations is evident. A
similar sequence is also observed for the asymptotic variable $\theta$,
as it is shown in Fig. \ref{fig1}(b). There are three small sequences of bifurcations, however we are not interested in their behavior 
here, but the same procedure can also be applied to them. Observe that, of course, the bifurcations
of the same period in (a) and (b) happen for the same value of the
control parameter $K$. As one can see in Fig. \ref{fig1} (c), as the parameter $K$ approaches the point of bifurcation, the Lyapunov 
exponent approaches zero. As discussed in \cite{ref15} the Lyapunov exponents are defined as
\begin{equation}
\lambda_j=\lim_{n\rightarrow\infty}{1\over{n}}\ln|\Lambda_j|~~,~~j=1,
2~~,
\label{eq4}
\end{equation}
where $\Lambda_j$ are the eigenvalues of
$M=\prod_{i=1}^nJ_i(I_i,\theta_i)$ and $J_i$ is the Jacobian matrix
evaluated over the orbit $(I_i,\theta_i)$. 
If at least one of the $\lambda_j$ is positive then the orbit is
classified as chaotic. We can see in Fig. (\ref{fig1}) (c) the behavior
of the Lyapunov exponents corresponding to both Figs.
(\ref{fig1})(a,b). It is also easy to see that when the bifurcations
happen, the exponent $\lambda$ vanishes. Such a behavior occurs because
the eigenvalues of the Jacobian matrix become complex numbers on unit circle. The Lyapunov exponents between $5.66<K<5.69$ correspond to the small sequence of bifurcation observed in Fig. \ref{fig1} (a,b) for the same range of the control parameter $K$.
It is important to emphasize that in some regions the Lyapunov exponent
assumes a constant and negative value.

\begin{figure}[t]
\centerline{\includegraphics[width=1.0\linewidth]{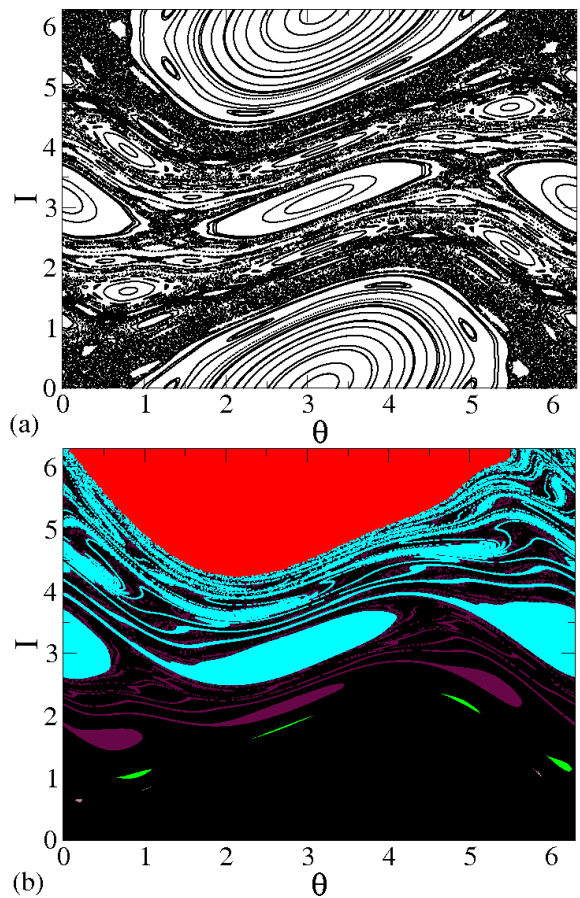}}
\caption{{(Color online) (a) Phase space for the conservative standard map $(\gamma=0)$ with $K=1$.
(b) Basin of attraction for the attracting fixed points (sinks) of period 1 (red and black), 2 (cyan), 3 (maroon)
and 4 (green). The control parameters used to
construct the basin of attraction were $K=1$ and $\gamma=10^{-2}$.}}
\label{fig11}
\end{figure}

\begin{figure}[t]
\centerline{\includegraphics[width=1.0\linewidth]{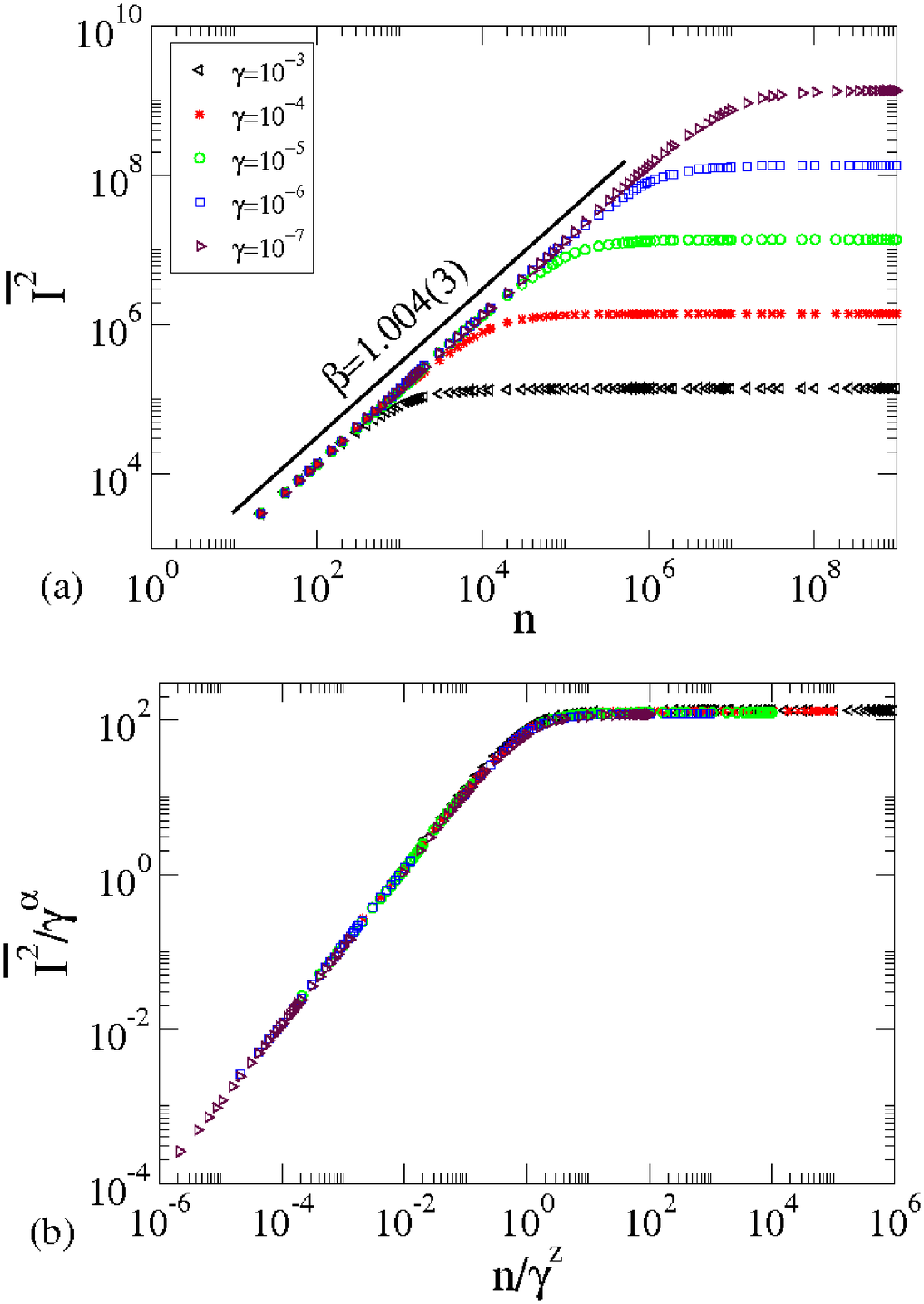}}
\caption{(Color online) (a) Behavior of $\overline{I^2}(n,\gamma)$ as function of $n$ for different values of the control parameter $\gamma$. 
(b) Their collapse onto a single universal plot.}
\label{fig2}
\end{figure}

\begin{figure}[t]
\centerline{\includegraphics[width=1.0\linewidth]{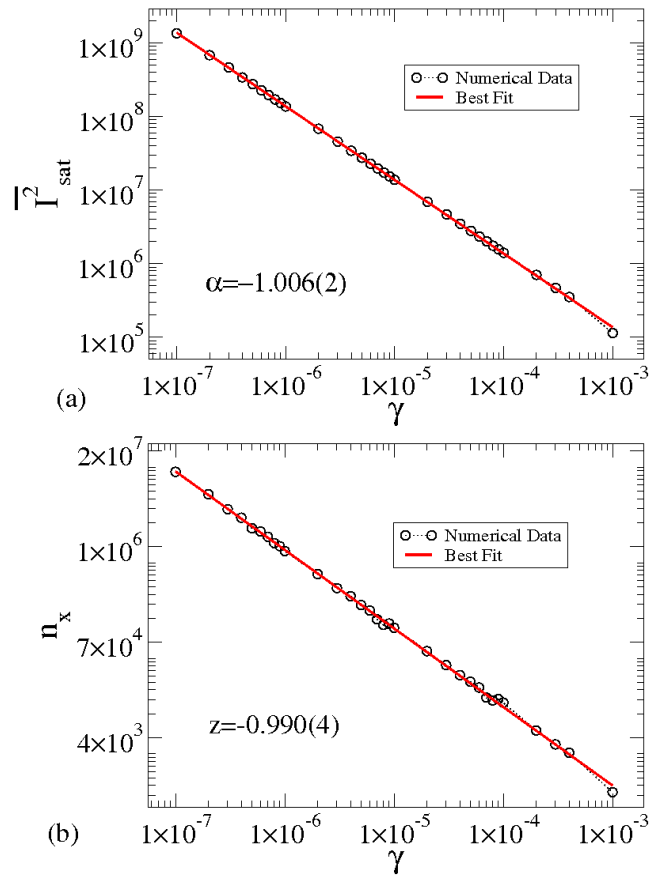}}
\caption{(Color online) (a) Plot of $\overline{I^2}_{sat}$ as function of the control parameter $\gamma$. (b) Behavior of the crossover number $n_x$ vs. $\gamma$.}
\label{fig3}
\end{figure}

The period doubling cascade was discovered by May \cite{ref15aa} in 1976 and by Grossmann and Thomae in 1977 \cite{ref15a} and later it was discovered by Feigenbaum \cite{ref16a,ref16b} 
that there is a universal feature along the bifurcations. 
The period doubling bifurcations converge geometrically to the chaos border. 
The procedure used to obtain the Feigenbaum constant $\delta$ is as follows: let $K_1$
represent the control parameter value at which period-1 gives birth to
a period-2 orbit, $K_2$ is the value where period-2 changes to
period-4 and so on. In general the parameter $K_n$ corresponds
to the control parameter value at which a period-$2^n$ orbit is born.
Thus, the Feigenbaum's $\delta$ is written as
\begin{equation}
\delta={\lim_{n\rightarrow \infty}{{K_{n}-K_{n-1}} \over
{K_{n+1}-K_{n}}}}.
\label{fei}
\end{equation}
The theoretical value for the Feigenbaum constant $\delta$ is $\delta=4.669201609\cdots$ . Considering the numerical data
obtained through the Lyapunov exponents calculation, the Feigenbaum's
$\delta$ obtained for the dissipative standard map is $\delta=4.6691(1)$. 
Since the numerical results are very hard to be obtained for bifurcations of 
higher orders, we have considered in our simulations only the bifurcations up to eleventh
order, but still our result is in a good agreement with the Feigenbaum's universal $\delta$.

\begin{figure}[t]
\centerline{\includegraphics[width=1.0\linewidth]{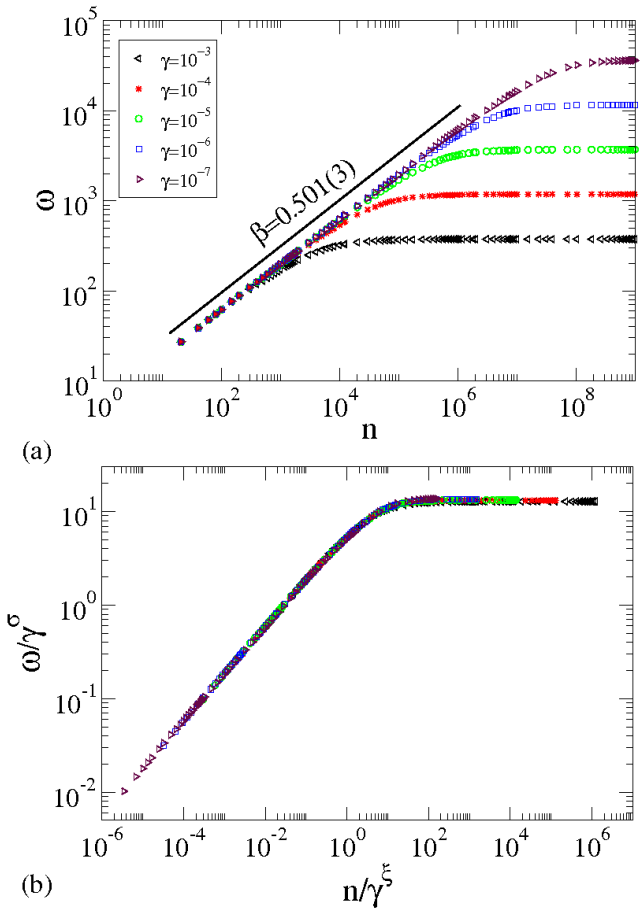}}
\caption{(Color online) (a) Behavior of $\omega(n,\gamma)$ as a function of $n$ for different values of the control parameter $\gamma$. 
(b) Their collapse onto a single universal plot.}
\label{fig4}
\end{figure}

\begin{figure}[t]
\centerline{\includegraphics[width=1.0\linewidth]{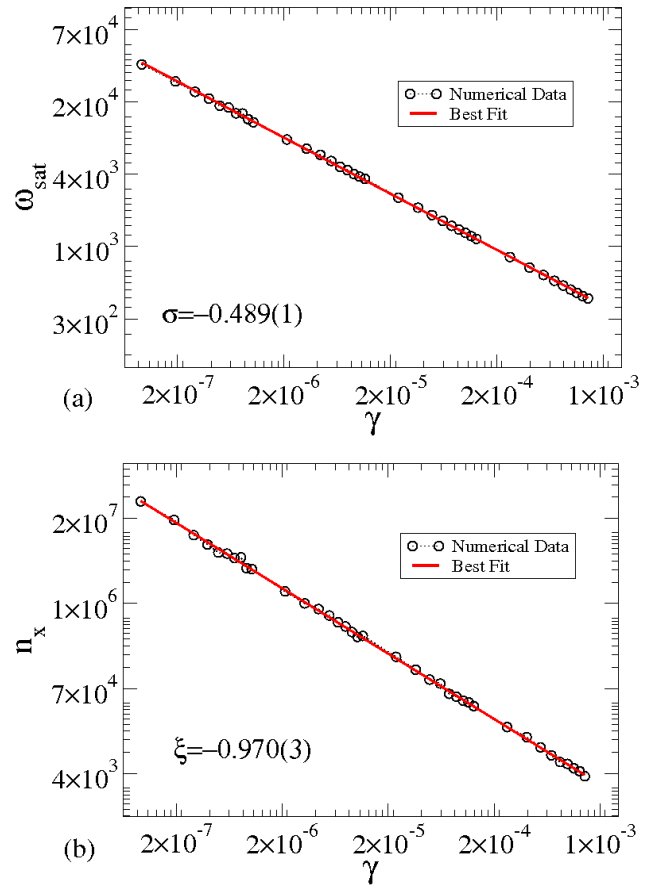}}
\caption{(Color online) (a) Plot of $\omega_{sat}$ as a function of $\gamma$. (b) Behavior of the crossover number $n_x$ vs. $\gamma$.}
\label{fig5}
\end{figure}

As we have shown for strong dissipation the system exhibits a sequence of period doubling bifurcations and for one of them we have found the 
Feigenbaum's $\delta$. Now, consider the parameter $K$ such that the conservative dynamics has the mixed phase space. 
When dissipation is taken into account the mixed structure is changed. Then, an elliptical fixed point (generally surrounded by KAM islands) turns into a sink. 
Regions of the chaotic sea might be replaced by chaotic attractors. 
In Fig. \ref{fig11}(a) the behavior of the phase space for the conservative dynamics $(\gamma=0)$ with $K=1$ is shown. As it is well known for
such a value of the parameter $K$ the last invariant torus is destroyed and the phase space has only KAM islands and a large chaotic sea.
Figure \ref{fig11}(b) shows the basin of attraction where the main fixed points are of period 1 (red and black), 2 (cyan), 3 (maroon) and 4 (green). 
The dissipation parameter is $\gamma=10^{-2}$. The procedure used to construct
the basin of attraction was to divide both $I$ and $\theta$ into grids of $1000$
parts each, thus leading to a total of $10^6$ different
initial conditions. Each initial condition was iterated up to $n=5
\times 10^5$. Apart from the mentioned periodic attractors there are many others, which, however, are difficult to find and display, because their basins of attraction are very small.
A more systematic overview of the dynamical phase diagram in the parameter-space $(K,\gamma)$, is given in reference \cite{ref27}.

When the control parameter $K$ is sufficiently large, all the regular regions are destroyed and the phase space for the conservative case is fully chaotic.
In this sense we will study some dynamical properties in the regime of $K>>K_c$ but taking into account weak dissipation. Our numerical results concern basically
the behavior of the $\overline{I^2}$ which is also the average energy $(\overline{E}=\overline{I^2}/2)$. Two steps were applied in order to obtain $\overline{I^2}$. Firstly, we
evaluate $\overline{I^2}$ over the orbit for a single initial
condition
\begin{eqnarray}
{I^2}_i(n,\gamma)={{1}\over{n+1}}\sum_{j=0}^nI^2_{i,j}~,
\label{eq17}
\end{eqnarray}
where the index $i$ corresponds to a sample of an ensemble of initial
conditions. Hence $\overline{I^2}$ is written as
\begin{eqnarray}
\overline{I^2}(n,\gamma)={{1}\over{M}}\sum_{i=1}^MI^2_i~.
\label{eq18}
\end{eqnarray}
We have iterated Eq. (\ref{eq18}) up to $n=10^{9}$ for an ensemble of $M=2000$ different initial conditions. 
The variable $I_0$ is fixed at $I_0=10^{-2}K$, $\theta_0$ is uniformly distributed on $\theta_0 \in [0,2\pi]$ and the control parameter was fixed as $K=20$. Figure \ref{fig2} 
shows the behavior of the $\overline{I^2}$ as a function of $n$ for five different values of the dissipation parameter $\gamma$, as labelled in the figure. 
It is easy to see in Fig. \ref{fig2} (a) two different kinds of behavior. For short $n$, $\overline{I^2}(n,\gamma)$ grows 
 according to a power law and suddenly it bends towards a regime of saturation for large enough values of $n$. The crossover from growth to the 
saturation is marked by a crossover iteration number $n_x$. This crossover number $n_x$ is actually quite well defined by the crossing of the two straight lines: The acceleration line and the saturation
line in the log-log plot.
Based on the behavior shown in Fig. \ref{fig2} (a), the following scaling hypotheses are assumed: 
\begin{enumerate}
\item{
When $n\ll{n_x}$, $\overline{I^2}(n,\gamma)$ grows according to
\begin{equation}
\overline{I^2}(n,\gamma)\propto {n}^{\beta}~,
\label{eq19}
\end{equation}
where the exponent $\beta$ is the acceleration exponent;
}
\item{
When $n\gg{n_x}$, $\overline{I^2}(n,\gamma)$ approaches a regime of saturation, which is described by
\begin{equation}
\overline{I^2}_{sat}(n ,\gamma)\propto \gamma^{\alpha}~,
\label{eq20}
\end{equation}
where the exponent $\alpha$ is the saturation exponent;
}
\item{
The crossover iteration number $n_x$ that marks the change from growth to the saturation is written as 
\begin{equation}
n_x\propto \gamma^{z}~,
\label{eq21}
\end{equation}
where $z$ is the crossover exponent.
}
\end{enumerate}

After considering these three initial suppositions, $\overline{I^2}(n,\gamma)$ can be described in terms of a scaling function of the type
\begin{equation}
\overline{I^2}(n,\gamma)=\tau \overline{I^2}(\tau^p{n},\tau^q{\gamma})~,
\label{eq22}
\end{equation}
where $\tau$ is the scaling factor, $p$ and $q$ are scaling exponents that in principle must be related to $\alpha$, $\beta$ and  $z$. 
Since $\tau$ is a scaling factor we can choose $\tau^p{n}=1$, or $\tau=n^{-1/p}$. Thus, Eq. (\ref{eq22}) is rewritten as
\begin{equation}
\overline{I^2}(n,\gamma)={n}^{-1/p}\overline{I^2}_1(n^{-q/p}\gamma)~,
\label{eq23}
\end{equation}
where the function $\overline{I^2}_1(n^{-q/p}\gamma)=\overline{I^2}(1,n^{-q/p}\gamma)$ is assumed to be constant for $n\ll{n_x}$. Comparing 
Eqs. (\ref{eq19}) and (\ref{eq23}), allows one to conclude that $\beta=-1/p$. Choosing now $\tau=\gamma^{-1/q}$ and Eq. (\ref{eq22}) is given by
\begin{equation}
\overline{I^2}(n,\gamma)=\gamma^{-1/q}\overline{I^2}_2(\gamma^{-p/q}n)~,
\label{eq24}
\end{equation}
where $\overline{I^2}_2(\gamma^{-p/q}n)=\overline{I^2}(\gamma^{-p/q} n,1)$. It is also assumed as constant for $n\gg{n_x}$. Comparison of Eq. (\ref{eq20}) and Eq. (\ref{eq24}) gives us $\alpha=-1/q$. 
Given the two different expressions of the scaling
factor $\tau$ in the crossover region $n \approx n_x$, we have $\tau=n^\beta_x=\gamma^\alpha$ and one must conclude that the crossover exponent $z$ is given by $z={{\alpha/\beta}}$. Observe that the 
scaling exponents are determined if the critical
exponents $\beta$ and $\alpha$ were numerically obtained. The exponent $\beta$ is obtained from a power law fitting for $\overline{I^2}(n,\gamma)$
curves for the parameter $\gamma\in[10^{-7},10^{-3}]$ for short iteration number, $n<<n_x$. Thus, the average of these values give
us $\beta=1.004(3)$.  Figure \ref{fig3} shows the behavior of (a) $\overline{I^2}_{sat} vs. \gamma$ and (b) $n_x vs. \gamma$. Applying 
power law fitting in the figure, it was found that $\alpha=-1.006(2)$ and $z=-0.990(4)$. Given $\alpha$ and $\beta$ the crossover exponent can also be obtained by solving 
$z={{\alpha/\beta}}$, yielding $z=-1.001(4)$. 
Such result indeed agrees with our numerical result. Finally, with the values of the critical exponents, the scaling hypothesis can be verified. 
Fig. \ref{fig2} (b) shows the collapse onto a single universal plot of 
five different curves of the $\overline{I^2}$ for different values of the dissipation parameter.

We have also considered the behavior of the average standard deviation of $I$, which is defined as 
\begin{equation}
\omega(n,\gamma)={{1}\over{M}}\sum_{i=1}^{M}\sqrt{\overline{I_i^2} (n,\gamma)-{\overline{I_i}}^2(n,\gamma)}~.
\label{eq18a}
\end{equation}
We have considered an ensemble of $M=2000$ different initial conditions iterated up to $10^{9}$. The variable $I_0$ has been fixed as $I_0=10^{-2}K$ and 
$\theta_0$ were uniformly distributed along $\theta_0 \in [0,2\pi]$.  Figure \ref{fig4} (a) shows the behavior of the $\omega$ as a function of the number of collisions $n$
for different values of the dissipation parameter and in Fig. \ref{fig4} (b) their collapse onto an universal plot. The acceleration exponent is $\beta=0.501(3)$.
In such a case $\omega_{sat} \propto {\gamma^{\sigma }}$ and 
$n_{x} \propto {\gamma^{\xi}} $, and $\sigma=-0.489(1)$ and $\xi=-0.970(3)$ as can be seen in Fig. \ref{fig5}. 
A similar scaling analysis shows a relationship $\xi=\sigma/\beta=-0.976(7)$, well in agreement with the numerical result for $\xi$.
This set of critical exponents
is the same as observed in a dissipative bouncer model \cite{andre}
where a classical particle collides inelastically with a time
periodically moving wall under the effect of gravity. It also is the
same set as observed for the oval-like billiard moving under the effect
of a quadratic frictional force \cite{buni}. Although both, the
systems and the kind of dissipation, are different, the phase transition
they are experiencing, namely, suppressing the unlimited growth of the
dynamical observable (energy for the oval-like billiard, deviation of
the average velocity for the dissipative bouncer and deviation of the
average action on the standard), belongs to the same class of
universality.

\begin{figure}[t]
\centerline{\includegraphics[width=1.0\linewidth]{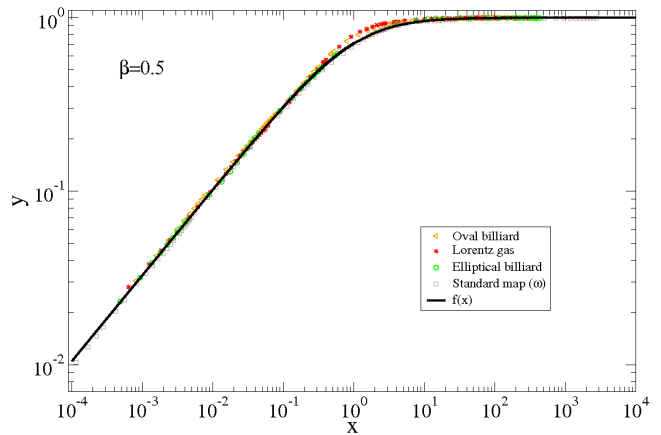}}
\caption{(Color online) Plot of $y$ as a function of $x$ for $\beta=0.5$.}
\label{fig6}
\end{figure}

\begin{figure}[t]
\centerline{\includegraphics[width=1.0\linewidth]{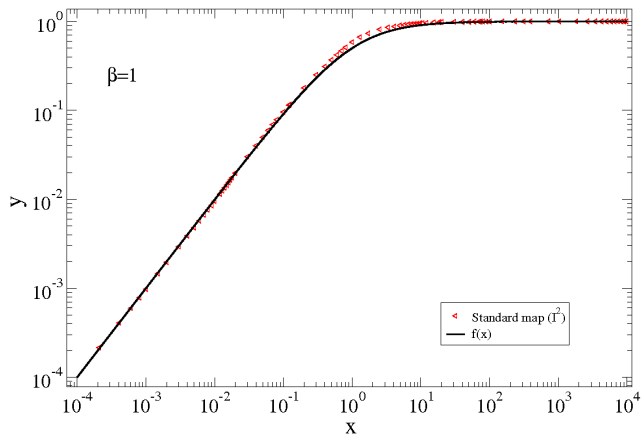}}
\caption{(Color online) Plot of $y=\overline{I^2}/\overline{I^2}_{sat}$ as a function of $x=n/n_x$ for the standard map where $\beta=1$.}
\label{fig7}
\end{figure}

The behavior observed in Fig. \ref{fig2} (b) and  Fig \ref{fig4} (b)
together with the
scaling discussion leads to an universal plot. This kind of behavior
was also observed in many different systems, under different approaches
for the conservative and dissipative dynamics and considering
different degrees of freedom mappings. To illustrate few applications
more than those discussed above in dissipative systems we nominate: (i)
oval billiard \cite{ref22}; (ii) Lorentz gas \cite{ref23} and; (iii)
elliptical billiard \cite{ref24}. For these cases, the transformation $\omega \rightarrow \omega/\omega_{sat}=y$ (or $\overline{V} \rightarrow \overline{V}/V_{sat}=y$ or $\overline{I^2} \rightarrow 
\overline{I^2}/\overline{I^2}_{sat}$)  and $n \rightarrow n/n_x=x$ 
all the curves for all these systems collapse onto a single universal plot as can be seen in Fig. \ref{fig6} (for $\beta=1/2$) and Fig. \ref{fig7} (for $\beta=1$). Based on this, we propose the following empirical function

\begin{equation}
y=f(x)={x^\beta \over {(1+x)^\beta}},
\label{eq20a}
\end{equation}
where the only parameter that remains is the acceleration exponent $\beta$. As one can see in Fig. \ref{fig6} and Fig. \ref{fig7} the agreement of our function $f(x)$ with the numerical data is quite 
good over many orders of magnitude with only some disagreement near the crossover. As mentioned, we have considered two values for the acceleration exponent, $\beta=0.5$ and $\beta=1$,
which includes a number of different physical systems.

\section{Conclusion}
\label{sec4}

As the concluding remark, some results for a dissipative standard map have been addressed. For the regime of strong
dissipation, the model exhibits a period doubling bifurcation cascade, where the so called Feigenbaum's
$\delta=4.6691(1)$ was numerically obtained. When weak dissipation is taken into account, for small values of $K$, the mixed structure
of the phase space is changed, the elliptic fixed points are replaced by attracting fixed points. No large chaotic attractors are observed in such a regime.
For large $K$, the behavior of the average energy is considered in the framework 
of scaling. Once the exponents  $\beta$, $\alpha$ and $z$ are obtained, the scaling hypotheses are confirmed by the perfect collapse of all curves onto a 
single universal plot. The  scaling behavior is also described in terms of the average standard deviation of the action variable. Given 
the exponents  $\beta$, $\sigma$ and $\xi$ we can confirm that the dissipative standard map belongs to the 
same class of universality of a dissipative time dependent elliptic billiard (2-D) \cite{ref24}, non-dissipative Fermi-Ulam model (1-D) \cite{ref25}, 
the corrugated wave guide (1-D) \cite{ref26}, for the range of the control parameters considered. Finally, we propose an empirical function $f(x)={x^\beta / {(1+x)^\beta}}$ in order to describe the 
universal scaling behavior observed when dissipation is taken into account for many different systems. This universal function gives a good description of the empirical 
numerical data over many orders of magnitude. Using it, only one empirical parameter is left in the system, namely the acceleration exponent $\beta$.

\acknowledgments
D.F.M.O. acknowledges the financial support by the Slovenian Human Resources Development and Scholarship Fund (Ad futura Foundation). 
M. R. acknowledges the financial support by The Slovenian Research Agency (ARRS). E. D. L. is grateful to FAPESP, CNPq and FUNDUNESP, Brazilian
agencies.


\begin{thebibliography}{10}

\bibitem{ref0} B. V. Chirikov, {\it Research concerning the theory of nonlinear resonance and stochasticity}, Preprint N 267, Institute of Nuclear Physics, Novosibirsk (1969).

\bibitem{ref1} B. V. Chirikov, Physics Reports, {\bf 52} (1979) 263.

\bibitem{ref2} G. M. Zaslavsky, {\it Hamiltonian Chaos and Fractional Dynamics} (Oxford,2006). 

\bibitem{ref3} A. J. Lichtenberg, M. A. Lieberman {\it { Appl. Math. Sci.}} {Vol 38} (Springer Verlag, New York,1992).

\bibitem{ref6} F. M. Izraelev, Physica D, {\bf  1}, (1980) 243.

\bibitem{ref8} T. H. Stix, Phys. Rev. Lett., {\bf  36} (1976) 10. 

\bibitem{ref4} H. L. Cycon, R. Froese, W. Kirsch, B. Simon {\it Sch\"ordinger Operators} (Berlin, Springer, 1987).

\bibitem{ref7} G. Casati, I. Guarneri, J. Ford, F. Vivaldi, Phys. Rev. A, {\bf  34} (1986) 1413.

\bibitem{ref7a} J. Martin, B. Georgeot, D. L. Shepelyansky, Phys. Rev. E {\bf 79} (2009) 066205.

\bibitem{ref5} S. Aubry, Physica D 7, (1983) 240.

\bibitem{ref9} M. Robnik,  {{ J. Phys. A: Math. Gen.}}, {\bf  16} (1983) 3971.

\bibitem{ref10} S. O. Kamphorst and S. P. Carvalho, { { Nonlinearity}}, {\bf  12} (1999) 1363.

\bibitem{ref11} V. Lopac, I. Mrkonji\'c and D. Radi\'c, {Phys. Rev. E}, {\bf  66} (2002) 036202.

\bibitem{ref12} V. Lopac, I. Mrkonji\'c, N. Pavin and D. Radi\'c, {{Physica D}}, {\bf  217} (2006) 88.

\bibitem{ref13} D. F. M. Oliveira  and E. D. Leonel,  { Commun Nonlinear Sci Numer Simulat} {\bf 15} (2010) 1092.

\bibitem{ref131} E. D. Leonel, A. L. P. Livorati, Physica. A, {\bf  387} (2008) 1155 . 

\bibitem{ref132} A. L. P. Livorati, D. G. Ladeira, E. D. Leonel, Physical Review E, {\bf 78} (2008) 056205. 

\bibitem{ref133} D. F. M.Oliveira, E. D. Leonel, Physics Letters A, {\bf  374} (2010) 3016.

\bibitem{ref1334} E. D. Leonel, P. V. E. McClintock,  J. Phys. A {\bf 38} (2005) L425.

\bibitem{ref13a} D. G. Ladeira, J. K. L. da Silva, Journal of Physics A, {\bf  40} (2007) 11467. 

\bibitem{ref14} G. M. Zaslavsky, ``The physics of chaos in Hamiltonian systems'' (Imperial College Press, London, (2007).

\bibitem{ref14a} G. M. Zaslavsky, Phys. Letts. A, {\bf  69A} (1978) 145.

\bibitem{ref15} J. P. Eckmann and D. Ruelle, Rev. Mod. Phys., {\bf {57}} (1985) 617.

\bibitem{ref15aa} R. M. May, Nature, {\bf 261} (1976) 459.

\bibitem{ref15a} S. Grossmann and S. Thomae, Z. Naturforsch., {\bf 32a} (1977) 1353.

\bibitem{ref16a} M. Feigenbaum, J. Stat. Phys., {\bf {19}} (1978) 25.

\bibitem{ref16b} M. Feigenbaum, J. Stat. Phys., {\bf {21}} (1979) 669.

\bibitem{ref27} D. F. M. Oliveira,  M. Robnik and E. D. Leonel, Submitted 2011.

\bibitem{andre} A. L. P. Livorati, D. G. Ladeira, and E. D. Leonel, Phys. Rev. E, {\bf 78}, (2008) 056205.

\bibitem{buni} E. D. Leonel and L. A. Bunimovich, Phys. Rev. E, {\bf 82} (2010) 016202.

\bibitem{ref22} D. F. M. Oliveira, E. D. Leonel,  Physica A, {\bf 389} (2010) 1009.

\bibitem{ref23} D. F. M. Oliveira, J. Vollmer, E. D. Leonel,  Physica D, {\bf  240} (2011) 389.

\bibitem{ref24} D. F. M. Oliveira and M. Robnik, Phys. Rev. E, {\bf 83} (2011) 26202.

\bibitem{ref25} E. D. Leonel, P. V. E. McClintock, J. k. L. Silva,  Phys. Rev. Lett., {\bf 93} (2004) 014101 . 

\bibitem{ref26} E. D. Leonel, Phys. Rev. Lett., {\bf 98} (2007) 114102 .



\end{thebibliography}
\end{document}